\documentclass[showpacs,preprintnumbers,twocolumn,eqsecnum,superscriptaddress,aps,nofootinbib]{revtex4}
\usepackage{amssymb}
\usepackage[centertags]{amsmath}
\usepackage{txfonts}
\usepackage{epsfig}
\usepackage{bm}
\usepackage{color}
\usepackage{graphicx,graphics}
\usepackage{multirow}
\usepackage{float}
\usepackage[pdfstartview=FitH]{hyperref}
\hypersetup{colorlinks=true, citecolor=blue, linkcolor=blue,filecolor=black,urlcolor=blue}
\allowdisplaybreaks[2]

\usepackage{slashed}
\usepackage{ulem}

\usepackage{booktabs}
\usepackage{multirow}

\begin{document}
\title{Three dimensional fragmentation functions from the quark-quark correlator}

\author{Kai-bao Chen}
\affiliation{School of Physics \& Key Laboratory of Particle Physics and Particle Irradiation (MOE), Shandong University, Jinan, Shandong 250100, China}

\author{Shu-yi Wei}
\affiliation{School of Physics \& Key Laboratory of Particle Physics and Particle Irradiation (MOE), Shandong University, Jinan, Shandong 250100, China}

\author{Wei-hua Yang}
\affiliation{School of Physics \& Key Laboratory of Particle Physics and Particle Irradiation (MOE), Shandong University, Jinan, Shandong 250100, China}

\author{Zuo-tang Liang}
\affiliation{School of Physics \& Key Laboratory of Particle Physics and Particle Irradiation (MOE), Shandong University, Jinan, Shandong 250100, China}

\begin{abstract}
We present the systematic results for three dimensional fragmentation functions 
defined via the quark-quark correlator for hadrons with spin 0, 1/2 and 1 respectively. 
These results are presented in terms of a spin independent part, a vector polarization dependent part 
and a tensor polarization dependent part. 
For spin 0 hadrons, only the spin independent part is needed, 
for spin 1/2 hadron, the polarization independent and vector polarization dependent parts are present, 
while for spin 1 hadrons, all the three parts exist. 
We discuss the general properties of these fragmentation functions and also present the 
corresponding results when transverse momentum is integrated.
\end{abstract}



\maketitle

\section{Introduction} \label{sec:I}
In describing high energy reactions, we need two sets of important quantities, the parton distribution functions (PDFs) 
and the fragmentation functions (FFs).
The former is used to describe the hadron structure and the latter describes the hadronization process. 
In a quantum field theoretical formulation, both PDFs and FFs are defined via the corresponding quark-quark correlator.
The quark-quark correlator is defined as a matrix depending on the hadron states. 
It is then decomposed into different components expressed in terms of the basic Lorentz covariants and the scalar functions. 
These scalar functions contain the information of the hadron structure and/or hadronization and are called the corresponding PDFs or FFs. 
In many cases in literature, specific components of the PDFs and/or FFs are introduced whenever needed,
sometimes with different conventions and/or notations. 
With the development of the related studies, it is necessary and useful to make a systematic study and present a complete set of such results. 
The results for three dimensional PDFs of the nucleon defined in this way are presented in \cite{Goeke:2005hb} in a systematical way. 
Since usually different types of hadrons with different flavors and spins are produced in a high energy reaction, 
FFs are therefore much more involved and perhaps even more interesting.  
Specific recent discussions can also be found e.g. in \cite{Collins:1992kk,Ji:1993vw, Mulders:1995dh,Boer:1997nt,Bacchetta:2000jk,Goeke:2003az,Bacchetta:2004zf,Bacchetta:2006tn,Liang:2006wp,Song:2010pf,Wei:2013csa,Wei:2014pma}.

The purpose of this note is to present the systematic results for three dimensional FFs for hadrons with spin 0, 1/2 and 1 obtained from 
the quark-quark correlator.  
We briefly describe the general procedure of deriving the results from the quark-quark correlator in Sec. \ref{sec:II}
and present the results for hadrons with different spins in Sec. \ref{sec:III}. 
We make a summary and a discussion in Sec. \ref{sec:IV}.

\section{General procedure of the decomposition} \label{sec:II}

Similar to that describes hadron structure, for quark fragmentation, the quark-quark correlator is defined as,
\begin{align}
\hat\Xi_{ij}^{(0)}(k_F;p,S) =\frac{1}{2\pi} & \sum_X \int  d^4\xi e^{-ik_F \xi} \langle 0| \mathcal{L}^\dag (0;\infty) \psi_i(0) |p,S;X\rangle \nonumber\\
& \times \langle p,S;X|\bar\psi_j(\xi) \mathcal{L}(\xi;\infty) |0\rangle, \label{qq-correlator}
\end{align}
where $k_F$ and $p$ denote the 4-momenta of the quark and that of the hadron respectively, $S$ denotes the spin of the hadron. 
$\mathcal{L}(\xi;\infty)$ is the gauge link that is given by \cite{Wei:2014pma},
\begin{align}
\mathcal{L}(\xi,&\infty) =Pe^{ig\int_{\xi^-}^\infty d\eta^-A^+(\eta_-;\xi^+,\vec{\xi}_\perp)} \nonumber\\
&= 1 + ig\int_{\xi^-}^\infty d \eta^- A^+(\eta^-;\xi^+,\vec{\xi}_\perp)) \nonumber\\
& + (ig)^2 \int_{\xi^-}^\infty d \eta_1^- \int_{\xi^-}^{\eta_1^-} d \eta_2^- A^+(\eta_2^-;\xi^+,\vec{\xi}_\perp)A^+(\eta_1^-;\xi^+,\vec{\xi}_\perp) \nonumber\\
& + \cdots.
\end{align}

It is clear that the correlator given by Eq.~(\ref{qq-correlator}) satisfies the following constraints 
imposed by hermiticity and parity conservation,  i.e., 
\begin{align}
& \hat\Xi^{\dagger(0)}(k_F;p,S) = \gamma^0 \hat\Xi^{(0)}(k_F;p,S) \gamma^0, \label{hermiticity}\\
& \hat\Xi^{(0)}(k_F;p,S) = \gamma^0 \hat\Xi^{(0)}(\tilde k_F; \tilde p, S^{\mathcal{P}}) \gamma^0, \label{parity}
\end{align}
where a tilded vector means $\tilde A^\mu = A_\mu$, and $S^{\mathcal{P}}$ means the polarization vector or tensor after space reflection. 
Unlike that for hadron structure, because of the presence of the gauge link and final state interactions between $h$ and $X$, 
time reversal puts no such simple constraint on the correlator $\hat\Xi^{(0)}(k_F;p,S)$.

The correlator $\hat\Xi^{(0)}(k_F;p,S)$ defined by Eq.~(\ref{qq-correlator}) depends on the 4-momentum $k_F$. 
The three dimensional FFs or the transverse momentum dependent (TMD) FFs are defined via 
the three dimensional quark-quark correlator $\hat\Xi^{(0)}(z,k_{F\perp};p,S)$ obtained from $\hat\Xi^{(0)}(k_{F},p,S)$ 
by integrating over $k_{F}^-$ , i.e. \cite{Wei:2014pma}, 
\begin{align}
\hat\Xi^{(0)}&(z,k_{F\perp};p,S) = \int \frac{p^+dk_F^+dk_F^-}{(2\pi)^2}2\pi\delta (k_F^+-p^+/z)\hat\Xi^{(0)}(k_{F},p,S)\nonumber\\
=&\sum_X \int \frac{p^+d\xi^-}{2\pi} d^2{\xi}_\perp e^{-i(p^+\xi^-/z - \vec{k}_{F\perp} \cdot \vec{\xi}_\perp)} \langle 0| \mathcal{L}^\dag (0;\infty) \psi(0) |p,S;X\rangle \nonumber\\
&~~~~~~~~~~~~\times  \langle p,S;X|\bar\psi(\xi) \mathcal{L}(\xi;\infty) |0\rangle, \label{Xi0}
\end{align}
where $z=p^+/k^+_F$ is the longitudinal momentum fraction defined in light cone coordinates. 
Here we use the light-cone coordinate and define the light-cone unit vectors as $\bar n = (1,0,\vec 0_\perp)$, $n = (0,1,\vec 0_\perp)$ and $n_\perp = (0,0,\vec 1)$. 
We choose the hadron's momentum as $z$-direction so that $p$ is decomposed as
$p^\mu = p^+ \bar n^\mu + \frac{M^2}{2p^+} n^\mu$.

The three dimensional or TMD FFs are obtained from $\hat\Xi^{(0)}(z,k_{F\perp};p,S)$ by decomposing it in the following steps. 
First, we note that $\hat\Xi^{(0)}(z,k_F;p,S)$ is a $4 \times 4$ matrix in Dirac indices and expand it in terms of the gamma matrices,
$\Gamma = \bigl\{ \mathbf{I},~ i\gamma_5,~ \gamma^\alpha,~ \gamma_5\gamma^\alpha,~ i\sigma^{\alpha\beta}\gamma_5 \bigr\}$, i.e.,
\begin{align}
\hat\Xi^{(0)}(z,k_{F\perp};&p,S) = \frac{1}{2} \Bigl[ \Xi^{(0)}(z,k_\perp;p,S) + i\gamma_5 \tilde\Xi^{(0)}(z,k_\perp;p,S) \nonumber\\
& + \gamma^\alpha \Xi_\alpha^{(0)}(z,k_\perp;p,S) + \gamma_5\gamma^\alpha \tilde\Xi_\alpha^{(0)}(z,k_\perp;p,S) \nonumber\\
& + i\sigma^{\alpha\beta}\gamma_5 \Xi_{\alpha\beta}^{(0)}(z,k_\perp;p,S) \Bigr]. \label{XiExpansion}
\end{align}
The coefficient functions are given by,
\begin{align}
\Xi&^{(0)[\Gamma]}(z,k_{F\perp};p,S) =  \frac{1}{2}{\rm{Tr}}\Bigl[\Gamma\hat\Xi^{(0)}(z,k_{F\perp};p,S)\Bigr]\nonumber\\
&=\frac{1}{2}\sum_X \int \frac{p^+d\xi^-}{2\pi} d^2\xi_\perp e^{-i(p^+\xi^-/z - \vec{k}_{F\perp} \cdot \vec{\xi}_\perp)}\nonumber\\
&\times 
 \langle p,S;X|\bar\psi(\xi) \mathcal{L}(\xi;\infty) |0\rangle \Gamma\langle 0|\mathcal{L}^\dag (0;\infty)\psi(0) |p,S;X\rangle, \label{eq:trace}
\end{align}
where $\Xi^{(0)[\Gamma]}$ represents respectively $\Xi^{(0)}$, $\tilde\Xi^{(0)}$,  $\Xi_\alpha^{(0)}$, $\tilde\Xi_\alpha^{(0)}$ and $\Xi_{\alpha\beta}^{(0)}$  
for different $\Gamma$'s.
Together with the demands imposed by the hermiticity and parity invariance [Eqs.~(\ref{hermiticity}) and (\ref{parity})],  
the Lorentz invariance demands that 
all the corresponding coefficient functions are real and are Lorentz scalar, pseudo-scalar, vector, axial-vector and tensor respectively.
Furthermore, the tensor $\Xi_{\alpha\beta}^{(0)}$  is anti-symmetric in Lorentz indices and odd under space reflection 
which implies that it can be made out of a vector and an axial-vector. 

Second, we expand these coefficient functions according to their respective Lorentz transformation properties 
in terms of the basic Lorentz covariants constructed from basic variables at hand. 
They are expressed as the sum of the basic Lorentz covariants multiplied by scalar functions of $z$ and $k_{F\perp}^2$. 
These scalar functions are the TMD FFs. 

Clearly, the basic Lorentz covariants that we can construct depend strongly on what basic variable(s) that we have at hand. 
Besides the 4-momenta $p$ and $k_F$, we have the variables describing the spin states.
Such variables are different for spin-0, 1/2 or 1 hadrons. 
We therefore obtain different results for hadrons with different spins. 

In the case of spin-$0$ hadron production, the independent variables that can be used are 
the 4-momenta $p_\alpha$ and $k_{F\perp\alpha}$, and the unit vector $n_\alpha$ that specifies the direction of the gauge link. 

For spin-1/2 hadrons, the polarization is described by a $2\times 2$ spin density matrix $\rho$
that is usually decomposed as $\rho = (1 + \vec S \cdot \vec \sigma)/2$, where $\vec\sigma$ is the Pauli matrix, 
and $\vec S$ is the polarization vector. It is usually represented by the covariant form $S=(0,\vec S)$ in the rest frame of the hadron. 
This means that in this case, besides $p_\alpha$, $k_{F\perp\alpha}$, and $n_\alpha$, we have an axial vector $S$ that can be used 
to construct the basic Lorentz covariants. 

For spin-1 hadrons, the polarization is described by a $3\times 3$ density matrix $\rho$. 
In this case, in the rest frame of the hadron, $\rho$ is usually decomposed as \cite{Bacchetta:2000jk},
\begin{align}
\rho = \frac{1}{3} (\mathbf{1} + \frac{3}{2}S^i \Sigma^i + 3 T^{ij} \Sigma^{ij}), \label{eq:spin1rho}
\end{align}
where, $\Sigma^i$ is the spin operator of spin-$1$ particle, 
and $\Sigma^{ij}= \frac{1}{2} (\Sigma^i\Sigma^j + \Sigma^j \Sigma^i) - \frac{2}{3} \mathbf{1} \delta^{ij}$. 
The spin polarization tensor $T^{ij}={\rm Tr}(\rho \Sigma^{ij})$, and is parameterized as,
\begin{align}
\mathbf{T}= \frac{1}{2}
\left(
\begin{array}{ccc}
-\frac{2}{3}S_{LL} + S_{TT}^{xx} & S_{TT}^{xy} & S_{LT}^x  \\
S_{TT}^{xy}  & -\frac{2}{3} S_{LL} - S_{TT}^{xx} & S_{LT}^{y} \\
S_{LT}^x & S_{LT}^{y} & \frac{4}{3} S_{LL}
\end{array}
\right).
\label{spintensor}
\end{align}
We see that, for spin-1 hadrons, besides the polarization vector $S$, we also need a tensor polarization part. 
The polarization vector is similar to that for spin-1/2 hadrons. 
The tensor polarization part has five independent components that are given by 
a Lorentz scalar $S_{LL}$, a Lorentz vector $S_{LT}^\mu = (0, S_{LT}^x, S_{LT}^y,0)$ 
and a Lorentz tensor $S_{TT}^{\mu\nu}$ that has two nonzero independent components 
$S_{TT}^{xx} = -S_{TT}^{yy}$ and $S_{TT}^{xy} = S_{TT}^{yx}$.
This means that for spin-1 hadrons, the quark-quark correlator $\hat\Xi^{(0)}$ can be written as the sum of 
a polarization independent part $\hat\Xi^{U(0)}$, a vector polarization dependent part $\hat\Xi^{V(0)}$ 
and a tensor polarization dependent part $\hat\Xi^{T(0)}$, i.e, 
\begin{align}
\hat\Xi^{(0)}(z,k_{F\perp};p,S) =& \hat\Xi^{U(0)}(z,k_{F\perp};p) + \hat\Xi^{V(0)}(z,k_{F\perp};p,S) \nonumber\\
+& \hat\Xi^{T(0)}(z,k_{F\perp};p,S).
\end{align}
Formally, the spin independent part is exactly the same as that for spin-0 hadrons, 
the vector polarization dependent part is the same as that for spin-1/2 hadrons. 
This means also that we only need to study each part separately. 
For spin-$0$ or unpolarized hadron production, we need then only the unpolarized part $\hat\Xi^{U(0)}(z,k_{F\perp};p)$, 
for spin-1/2 hadrons, both the spin independent part $\hat\Xi^{U(0)}(z,k_{F\perp};p)$ and 
the vector polarization dependent part $\hat\Xi^{V(0)}(z,k_{F\perp};p,S)$ contribute. 
The tensor polarization dependent part $\hat\Xi^{T(0)}(z,k_{F\perp};p,S)$ contributes only in the case of spin-$1$ hadron production.
We present the results in next section.

\section{TMD FFs for hadrons with different spins} \label{sec:III}
In this section, we present the results for the decomposition of the quark-quark correlator Eq.(\ref{Xi0}).

\subsection{Spin-$0$} \label{FFs-Spin0}
In the case of spin-$0$ hadron production, only the spin independent part $\hat\Xi^{U(0)}(z,k_{F\perp};p)$ contributes. 
The independent variables that can be used to construct the basic Lorentz covariants are $p_\alpha$, $k_{F\perp\alpha}$, and $n_\alpha$. 
The basic Lorentz convariants that we can construct from them are: one Lorentz scalar $p^2=M^2$, no pseudo-scalar,  
three Lorentz vectors, $p$, $k_{F\perp}$ and $n$, one axial vector $\varepsilon_{\perp\alpha\beta}k_{F\perp}^\beta$, 
and three anti-symmetric and space reflection odd Lorentz tensors $p_{[\rho}\varepsilon_{\perp\alpha]\beta}k_{F\perp}^\beta$, 
$\varepsilon_{\perp\rho\alpha}$ and $n_{[\rho}\varepsilon_{\perp\alpha]\beta}k_{F\perp}^\beta$.
Hence, the general decomposition of the spin independent part of the quark-quark correlator is given by,
\begin{align}
& z\Xi^{U(0)}(z,k_{F\perp};p) = ME(z,k_{F\perp}), \label{eq:XiUS}\\
& z\tilde\Xi^{U(0)}(z,k_{F\perp};p) =0, \label{eq:XiPS}\\
& z\Xi_\alpha^{U(0)}(z,k_{F\perp};p) = p^+ \bar n_\alpha D_1(z,k_{F\perp})+ k_{F\perp\alpha} D^\perp(z,k_{F\perp}) \nonumber\\
& \hspace{2.6cm} + \frac{M^2}{p^+}n_\alpha D_3(z,k_{F\perp}),\label{eq:XiUV}\\
 & z\tilde\Xi_\alpha^{U(0)}(z,k_{F\perp};p) = \varepsilon_{\perp\alpha\beta}k_{F\perp}^\beta G^\perp(z,k_{F\perp}), \label{eq:XiUAV}\\
& z\Xi_{\rho\alpha}^{U(0)}(z,k_{F\perp} ;p) = \frac{p^+ \bar n_{[\rho}\varepsilon_{\perp\alpha]\beta}k_{F\perp}^\beta}{M} H_1^\perp(z,k_{F\perp}) + M\varepsilon_{\perp\rho\alpha} H(z,k_{F\perp}) \nonumber\\
& \hspace{2.6cm} + \frac{M}{p^+} n_{[\rho}\varepsilon_{\perp\alpha]\beta}k_{F\perp}^\beta H_3^\perp(z,k_{F\perp}), \label{eq:XiUT}
\end{align}
where $\varepsilon_{\perp\rho\sigma} \equiv \varepsilon_{\alpha\beta\rho\sigma} \bar n^\alpha n ^\beta$, and the commutation symbol $A^{[\rho}B^{\sigma]} \equiv A^\rho B^\sigma - A^\sigma B^\rho$.
Here, we note in particular that, compared with the corresponding $\bar n$ component, 
the $n_\perp$ and $n$ components are suppressed by $M/p^+$ and $(M/p^+)^2$ 
and contribute at twist-3 and twist-4 respectively.
We see that there are 8 unpolarized TMD FFs, 2 of them contribute at twist-2, 4 at twist-3 and the other 2 at twist-4 level. 

The notation used here takes the following convention: 
$D$, $G$ and $H$ are for unpolarized, longitudinally polarized and transversely polarized quarks, those defined via the scalar and pseudo-scalar are denoted by $E$;
a number in the subscripts specifies the twist: $1$ for twist-2, no number for twist-3 and a 3 for twist-4; 
we will also use different symbols in the subscripts to denote the polarization of the produced hadron such $L$ and $T$   
in the vector polarization case and $LL$, $LT$ or $TT$ in the tensor polarization case; 
a $\perp$ in the superscript denotes that the corresponding basic Lorentz covariant is $k_{F\perp}$-dependent.

If we decompose the quark field in Eq.~(\ref{eq:trace}) into the sum of the right-
and left-handed parts, i.e., $\psi = \psi_R + \psi_L$ with $\psi_{R/L} \equiv \frac{1}{2}(1\pm\gamma_5)\psi$.
We see that for $\Gamma = \mathbf{I}$, $i\gamma_5$ and $i\sigma^{\alpha\beta}\gamma_5$, 
$\bar\psi_R\Gamma\psi_L$ and $\bar\psi_L\Gamma\psi_R$ are non-zero.
So the terms related to them (i.e., the $E$'s and the $H$'s) 
correspond to helicity-flipped quark structure and are called chiral-odd ($\chi$-odd). 
Similarly, for $\Gamma = \gamma^\alpha$ and $\gamma^5\gamma^\alpha$, 
$\bar\psi_L\Gamma\psi_L$ and $\bar\psi_R\Gamma\psi_R$ are non-zero.
Hence, the terms related to them (i.e. the $D$'s and the $G$'s) do not flip the quark helicity and are $\chi$-even.
We also recall the properties of the fermion bilinears 
under time-reversal $\hat{\mathcal{T}}$, i.e., 
\begin{align}
\hat{\mathcal{T}}&\Bigl\{ \bar\psi \psi,~ \bar\psi i\gamma_5 \psi,~ \bar\psi \gamma_\alpha \psi, \bar\psi \gamma_5\gamma_\alpha\psi,~ \bar\psi i\sigma_{\alpha\beta}\gamma_5 \psi  \Bigr\}\nonumber\\
\Rightarrow 
&\Bigl\{ \bar\psi \psi,~ -\bar\psi i\gamma_5 \psi,~ \bar\psi \gamma^\alpha \psi,~ \bar\psi \gamma_5\gamma^\alpha\psi,~ \bar\psi i\sigma^{\alpha\beta}\gamma_5 \psi  \Bigr\}.
\end{align}
Using this, we can determine whether a component of FF defined via quark-quark correlator given by Eqs.~(\ref{eq:XiUS}-\ref{eq:XiUT}) 
is time reversal even (T-even) or odd (T-odd) according to the time reversal behavior of the corresponding basic Lorentz covariant. 
In this way, we find out that $G^\perp$, $H_1^\perp$, $H$ and $H_3^\perp$ are T-odd, all the others in Eqs.~(\ref{eq:XiUS}-\ref{eq:XiUT}) are T-even.
We also note that they are usually referred as ``naive T-odd'' or ``naive T-even'' because the interactions between the produced hadron $h$ and the 
rest $X$ can destroy simple regularities so all of them can exist in a practical hadronization process. 

If we integrate over $d^2k_{F\perp}$, terms with $k_{F\perp}$ odd Lorentz structures vanish. We obtain,
\begin{align}
& z\Xi^{U(0)}(z;p) = ME(z), \\
 & z\tilde\Xi^{U(0)}(z;p) =0, \\
& z\Xi_\alpha^{U(0)}(z;p) = p^+ \bar n_\alpha D_1(z) + \frac{M^2}{p^+}n_\alpha D_3(z),\\
 & z\tilde\Xi_\alpha^{U(0)}(z;p) = 0, \\
& z\Xi_{T\rho\alpha}^{U(0)}(z;p) = M\varepsilon_{\perp\rho\alpha} H(z),
\end{align}
where the one dimensional FF is just equal to the corresponding three dimensional one integrated over $d^2k_{F\perp}$. 
We see that there are only 4 left and the number density $D_1(z)$ is the only leading twist, 
2 of them contribute at twist-3 and the other one at twist-4.

\subsection{Spin-$1/2$} \label{FFs-Spin1/2}
For spin-$1/2$ hadrons, the vector polarization dependent part contributes. 
We have, besides $p_\alpha$, $k_{F\perp\alpha}$, and $n_\alpha$, the polarization vector $S$ to use to construct the basic Lorentz covariants. 
The polarization vector $S$ is decomposed as,
\begin{align}
S^\mu = \lambda \frac{p^+}{M} \bar n^\mu + S_T^\mu - \lambda \frac{M}{2p^+} n^\mu,
\end{align}
where $\lambda$ denotes the helicity of the hadron and $S_T = (0,0,\vec S_T)$ denotes the transverse polarization.

We build the $S$-dependent basic Lorentz covariants with the corresponding properties under space reflection as demanded 
and obtain the general decomposition of the $S$-dependent part of the quark-quark correlator as,
\begin{widetext}
\begin{align}
& z\Xi^{V(0)}(z,k_{F\perp};p,S) = \varepsilon_{\perp\rho\sigma}k_{F\perp}^\rho S_T^\sigma E_T^\perp(z,k_{F\perp}), \\
& z\tilde \Xi^{V(0)}(z,k_{F\perp};p,S) =M \Bigl[ \lambda E_L(z,k_{F\perp}) + \frac{k_{F\perp} \cdot S_T}{M} E_T(z,k_{F\perp}) \Bigr], \\
& z\Xi_\alpha^{V(0)}(z,k_{F\perp};p,S) = p^+ \bar n_\alpha \frac{\varepsilon_{\perp\rho\sigma}k_{F\perp}^\rho S_T^\sigma}{M}D_{1T}^\perp(z,k_{F\perp}) + M\varepsilon_{\perp\alpha\rho}S_T^\rho D_T(z,k_{F\perp}), \nonumber\\
& \hspace{2cm} + \varepsilon_{\perp\alpha\rho}k_{F\perp}^\rho \Bigl[ \lambda D_L^\perp(z,k_{F\perp}) + \frac{k_{F\perp} \cdot S_T}{M}D_T^{\perp}(z,k_{F\perp}) \Bigr] + \frac{M}{p^+}n_\alpha \varepsilon_{\perp\rho\sigma}k_{F\perp}^\rho S_T^\sigma D_{3T}^\perp(z,k_{F\perp}), \\
& z\tilde\Xi_\alpha^{V(0)}(z,k_{F\perp};p,S) = p^+ \bar n_\alpha \Bigl[ \lambda G_{1L}(z,k_{F\perp}) + \frac{k_{F\perp} \cdot S_T}{M} G_{1T}^\perp(z,k_{F\perp}) \Bigr] + MS_{T\alpha} G_T(z,k_{F\perp}) \nonumber \\
& \hspace{2cm} + k_{F\perp\alpha} \Bigl[ \lambda G_{L}^\perp(z,k_{F\perp}) + \frac{k_{F\perp} \cdot S_T}{M} G_{T}^{\perp}(z,k_{F\perp}) \Bigr] + \frac{M^2}{p^+}n_\alpha \Bigl[ \lambda G_{3L}(z,k_{F\perp}) + \frac{k_{F\perp} \cdot S_T}{M} G_{3T}(z,k_{F\perp}) \Bigr], \\
& z\Xi_{\rho\alpha}^{V(0)}(z,k_{F\perp};p,S) = p^+ \bar n_{[\rho}S_{T\alpha]} H_{1T}(z,k_{F\perp}) + \frac{p^+ \bar n_{[\rho}k_{F\perp\alpha]}}{M} \Bigl[ \lambda H_{1L}^\perp(z,k_{F\perp}) + \frac{k_{F\perp} \cdot S_T}{M} H_{1T}^\perp(z,k_{F\perp}) \Bigr] \nonumber\\
& \hspace{2cm} + k_{F\perp[\rho}S_{T\alpha]} H_T^\perp(z,k_{F\perp}) + \bar n_{[\rho}n_{\alpha]} \Bigl[ M\lambda H_L(z,k_{F\perp}) + k_{F\perp} \cdot S_TH_T^{\prime\perp}(z,k_{F\perp}) \Bigr] \nonumber\\
& \hspace{2cm} + \frac{M^2}{p^+} \Bigl\{ n_{[\rho}S_{T\alpha]} H_{3T}(z,k_{F\perp}) + \frac{n_{[\rho}k_{F\perp\alpha]}}{M} \Bigl[ \lambda H_{3L}^\perp(z,k_{F\perp}) + \frac{k_{F\perp} \cdot S_T}{M} H_{3T}^\perp(z,k_{F\perp}) \Bigr] \Bigr\}.
\end{align}
We see that there are 24 vector polarization dependent TMD FFs, 6 of them contribute at twist-2, 12 at twist-3 and the other 6 at twist-4 level. 
Among them, 8 are naive T-odd ($E_T^\perp$, $E_L$, $E_T$, $D_{1T}^\perp$, $D_L^\perp$, $D_T$, $D_T^{\perp}$ and $D_{3T}^\perp$), 
and the other 16 are T-even.

If we integrate over $d^2k_{F\perp}$, only 8 survive, i.e.,
\begin{align}
& z\Xi^{V(0)}(z;p,S) = 0, \\
& z\tilde \Xi^{V(0)}(z;p,S) = \lambda M E_L(z), \\
& z\Xi_\alpha^{V(0)}(z;p,S) = M \varepsilon_{\perp\alpha\rho} S_T^\rho D_{T}(z), \\
& z\tilde\Xi_\alpha^{V(0)}(z;p,S) = \lambda p^+ \bar n_\alpha  G_{1L}(z)  + MS_{T\alpha} G_T(z) + \lambda \frac{M^2}{p^+}n_\alpha  G_{3L}(z), \\
& z\Xi_{T\rho\alpha}^{V(0)}(z;p,S) = p^+ \bar n_{[\rho}S_{T\alpha]} H_{1T}(z) + \lambda M \bar n_{[\rho}n_{\alpha]}  H_L(z) + \frac{M^2}{p^+}  n_{[\rho}S_{T\alpha]} H_{3T}(z),
\end{align}
where the one dimensional FF is just equal to the corresponding three dimensional one integrated over $d^2k_{F\perp}$ with the following four 
exceptions,  
\begin{align}
& D_{T}(z) = \int \frac{d^2k_{F\perp}}{(2\pi)^2} \Bigl( D_T(z,k_{F\perp}) + \frac{k_{F\perp}^2}{2M^2} D_T^{\perp}(z,k_{F\perp}) \Bigr), \\
& G_T(z) = \int \frac{d^2k_{F\perp}}{(2\pi)^2} \Bigl( G_T(z,k_{F\perp}) + \frac{k_{F\perp}^2}{2M^2} G_T^{\perp}(z,k_{F\perp}) \Bigr),\\
& H_{1T}(z) = \int \frac{d^2k_{F\perp}}{(2\pi)^2} \Bigl( H_{1T}(z,k_{F\perp}) + \frac{k_{F\perp}^2}{2M^2} H_{1T}^{\perp}(z,k_{F\perp}) \Bigr),\\
& H_{3T}(z) = \int \frac{d^2k_{F\perp}}{(2\pi)^2} \Bigl( H_{3T}(z,k_{F\perp}) + \frac{k_{F\perp}^2}{2M^2} H_{3T}^{\perp}(z,k_{F\perp}) \Bigr).
\end{align}
We see that, in the one dimensional case, for the vector polarization dependent part, we have totally 2 leading twist FFs, 
they are the longitudinal spin transfer $G_{1L}(z)$ and the transverse spin transfer $H_{1T}(z)$. 
We also have 4 twist-3 that lead to induced polarization of hadron and 2 twist-4 FFs that are addenda to the 
longitudinal and transverse spin transfer respectively.

\subsection{Spin-$1$} \label{FFs-Spin1}
For spin-1 hadrons, we have the spin independent part, the vector polarization dependent part and the tensor polarization dependent part.
The spin independent and vector polarization dependent parts take exactly the same forms as those presented in Sec.\ref{FFs-Spin0} and \ref{FFs-Spin1/2}.
Here, we present the tensor polarization dependent part only.

To obtain the tensor polarization dependent part, we construct basic Lorentz covariants by using, 
besides $p_\alpha$, $k_{F\perp\alpha}$ and $n_\alpha$, the Lorentz scalar $S_{LL}$, Lorentz vector $S_{LT}$, 
and Lorentz tensor $S_{TT}$. 
We note that the tensor polarization $T^{\mu\nu}$ is expressed in terms of $S_{LL}$, $S_{LT}^\mu$ and $S_{TT}^{\mu\nu}$ in the following way, i.e., 
\begin{align}
T^{\mu\nu} = \frac{1}{2} \Bigl[ \frac{4}{3}S_{LL}\Bigl( \frac{p^+}{M} \Bigr)^2 \bar n^\mu \bar n^\nu + \frac{p^+}{M} n^{\{\mu}S_{LT}^{\nu\}} - \frac{2}{3}S_{LL}(\bar n^{\{\mu}n^{\nu\}} - g_\perp^{\mu\nu}) + S_{TT}^{\mu\nu} - \frac{M}{2p^+} \bar n^{\{\mu}S_{LT}^{\nu\}} + \frac{1}{3}S_{LL}\Bigl( \frac{M}{p^+} \Bigr)^2 n^\mu n^\nu \Bigr],
\end{align}
where we used the anti-commutation symbol $A^{\{\mu}B^{\nu\}} \equiv A^\mu B^\nu + A^\nu B^\mu$, and $g_\perp^{\mu\nu} \equiv g^{\mu\nu} - \bar n^\mu n^\nu - n^{\mu} \bar n^\nu$. 
This is useful to specify the twists of the corresponding terms. 
In this way, we obtain the most general decomposition for the tensor polarization dependent part as given by,
\begin{align}
& z\Xi^{T(0)}(z,k_{F\perp};p,S) =M \Bigl[ S_{LL}E_{LL}(z,k_{F\perp}) + \frac{k_{F\perp} \cdot S_{LT}}{M} E_{LT}^\perp(z,k_{F\perp}) + \frac{k_{F\perp} \cdot S_{TT} \cdot k_{F\perp}}{M^2} E_{TT}^{\perp}(z,k_{F\perp})\Bigr], \\
& z\tilde \Xi^{T(0)}(z,k_{F\perp};p,S) =M \Bigl[ \frac{\epsilon_\perp^{k_F S_{LT}}}{M} E_{LT}^{\prime\perp} (z,k_{F\perp}) + \frac{\epsilon_{\perp k_F \alpha} k_\beta S_{TT}^{\alpha\beta}}{M^2} E_{TT}^{\prime\perp}(z,k_{F\perp}) \Bigr], \\
& z\Xi_\alpha^{T(0)}(z,k_{F\perp};p,S) = p^+ \bar n_\alpha \Bigl[ S_{LL} D_{1LL}(z,k_{F\perp}) + \frac{k_{F\perp} \cdot S_{LT}}{M}D_{1LT}^\perp (z,k_{F\perp}) + \frac{k_{F\perp} \cdot S_{TT} \cdot k_{F\perp}}{M^2} D_{1TT}^\perp(z,k_{F\perp}) \Bigr] \nonumber\\
 & \hspace{2cm} + k_{F\perp\alpha} \Bigl[ S_{LL}D_{LL}(z,k_{F\perp}) + \frac{k_{F\perp} \cdot S_{LT}}{M}D_{LT}^\perp (z,k_{F\perp}) + \frac{k_{F\perp} \cdot S_{TT} \cdot k_{F\perp}}{M^2} D_{TT}^\perp(z,k_{F\perp}) \Bigr] \nonumber\\
 & \hspace{2cm} + M S_{LT\alpha} D_{LT}(z,k_{F\perp}) + k_{F\perp}^\rho S_{TT\rho\alpha} D_{TT}^{\prime\perp}(z,k_{F\perp}) \nonumber\\
 & \hspace{2cm} + \frac{M^2}{p^+}n_\alpha \Bigl[ S_{LL} D_{3LL}(z,k_{F\perp}) + \frac{k_{F\perp} \cdot S_{LT}}{M}D_{3LT}^\perp (z,k_{F\perp}) + \frac{k_{F\perp} \cdot S_{TT} \cdot k_{F\perp}}{M^2} D_{3TT}^\perp(z,k_{F\perp}) \Bigr], \\
& z\tilde \Xi_\alpha^{T(0)}(z,k_{F\perp};p,S) = p^+ \bar n_\alpha \Bigl[ \frac{\varepsilon_\perp^{k_{F\perp}S_{LT}}}{M}G_{1LT}^\perp(z,k_{F\perp}) + \frac{\varepsilon_{\perp k_{F\perp}\rho}k_{F\perp\sigma}S_{TT}^{\rho\sigma}}{M^2}G_{1TT}^\perp(z,k_{F\perp}) \Bigr] \nonumber\\
 & \hspace{2cm} + \varepsilon_{\perp\alpha\rho}k_{F\perp}^\rho \Bigl[ S_{LL}G_{LL}^\perp + \frac{k_{F\perp} \cdot S_{LT}}{M}G_{LT}^\perp (z,k_{F\perp}) + \frac{k_{F\perp} \cdot S_{TT} \cdot k_{F\perp}}{M^2} G_{TT}^\perp(z,k_{F\perp}) \Bigr] \nonumber\\
 & \hspace{2cm} + M\varepsilon_{\perp\alpha\rho}S_{LT}^\rho G_{LT}(z,k_{F\perp}) + \varepsilon_{\perp\alpha\rho}k_{F\perp\sigma}S_{TT}^{\rho\sigma} G_{TT}^{\prime\perp}(z,k_{F\perp}) \nonumber\\
 & \hspace{2cm} + \frac{M^2}{p^+}n_\alpha \Bigl[ \frac{\varepsilon_\perp^{k_{F\perp}S_{LT}}}{M}G_{3LT}^\perp(z,k_{F\perp}) + \frac{\varepsilon_{\perp k_{F\perp}\rho}k_{F\perp\sigma}S_{TT}^{\rho\sigma}}{M^2}G_{3TT}^\perp(z,k_{F\perp}) \Bigr]  ,\\
& z\Xi_{\rho\alpha}^{T(0)}(z,k_{F\perp} ;p,S) = \frac{p^+ \bar n_{[\rho} \varepsilon_{\perp\alpha]\sigma}k_{F\perp}^\sigma}{M} \Bigl[ S_{LL} H_{1LL}^\perp(z,k_{F\perp}) + \frac{k_{F\perp} \cdot S_{LT}}{M}H_{1LT}^\perp (z,k_{F\perp}) + \frac{k_{F\perp} \cdot S_{TT} \cdot k_{F\perp}}{M^2} H_{1TT}^\perp(z,k_{F\perp}) \Bigr] \nonumber\\
 & \hspace{2cm} + p^+ \bar n_{[\rho} \varepsilon_{\perp\alpha]\sigma}S_{LT}^\sigma H_{1LT}(z,k_{F\perp}) + \frac{p^+ \bar n_{[\rho} \varepsilon_{\perp\alpha]\sigma}k_{F\perp\delta}S_{TT}^{\sigma\delta}}{M} H_{1TT}^{\prime\perp}(z,k_{F\perp}) \nonumber\\
 & \hspace{2cm} + M\varepsilon_{\perp\rho\alpha} \Bigl[ S_{LL} H_{LL}(z,k_{F\perp}) + \frac{k_{F\perp} \cdot S_{LT}}{M}H_{LT}^\perp (z,k_{F\perp}) + \frac{k_{F\perp} \cdot S_{TT} \cdot k_{F\perp}}{M^2} H_{TT}^\perp(z,k_{F\perp}) \Bigr] \nonumber\\
 & \hspace{2cm} + \bar n_{[\rho}n_{\alpha]} \Bigl[ \varepsilon_\perp^{k_{F\perp}S_{LT}} H_{LT}^{\prime\perp}(z,k_{F\perp}) + \frac{\varepsilon_{\perp k_{F\perp}\sigma}k_{F\perp\delta}S_{TT}^{\sigma\delta}}{M}H_{TT}^{\prime\perp}(z,k_{F\perp}) \Bigr] \nonumber\\
 & \hspace{2cm} + \frac{M^2}{p^+} \Bigl\{ \frac{n_{[\rho} \varepsilon_{\perp\alpha]\sigma}k_{F\perp}^\sigma}{M} \Bigl[ S_{LL} H_{3LL}^\perp(z,k_{F\perp}) + \frac{k_{F\perp} \cdot S_{LT}}{M}H_{3LT}^\perp (z,k_{F\perp}) + \frac{k_{F\perp} \cdot S_{TT} \cdot k_{F\perp}}{M^2} H_{3TT}^\perp(z,k_{F\perp}) \Bigr] \nonumber\\
 & \hspace{2cm} \quad + n_{[\rho} \varepsilon_{\perp\alpha]\sigma}S_{LT}^\sigma H_{3LT}(z,k_{F\perp}) + \frac{n_{[\rho} \varepsilon_{\perp\alpha]\sigma}k_{F\perp\delta}S_{TT}^{\sigma\delta}}{M} H_{3TT}^{\prime\perp}(z,k_{F\perp}) \Bigr\}.
\end{align}
We see that there are totally 40 tensor polarization dependent TMD FFs, 
10 contribute at twist-2, 20 at twist-3 and the other 10 at twist-4. 
Among them, 24 (those related to $\tilde \Xi_\alpha^{T(0)}$ and $\Xi_{\rho\alpha}^{T(0)}$) are T-odd and the other 16 are T-even. 

We integrate over $d^2k_{F\perp}$ and obtain, 
\begin{align}
& z\Xi^{T(0)}(z;p,S) =M S_{LL}E_{LL}(z), \\
& z\tilde\Xi^{T(0)}(z;p,S) = 0, \\
& z\Xi_\alpha^{T(0)}(z;p,S) = p^+ \bar n_\alpha S_{LL} D_{1LL}(z) + MS_{LT\alpha} D_{LT}(z,k_{F\perp}) +   \frac{M^2}{p^+}n_\alpha S_{LL} D_{3LL}(z), \\
& z\tilde\Xi_\alpha^{T(0)}(z;p,S) = M \varepsilon_{\perp\alpha\rho} S_{LT}^\rho G_{LT} (z),\\
& z\Xi_{\rho\alpha}^{T(0)}(z ;p,S) = p^+ \bar n_{[\rho} \varepsilon_{\perp\alpha]\sigma}S_{LT}^\sigma H_{1LT}(z) 
  + M\varepsilon_{\perp\rho\alpha} S_{LL} H_{LL}(z)
  + \frac{M^2}{p^+}  n_{[\rho} \varepsilon_{\perp\alpha]\sigma} S_{LT}^\sigma H_{3LT}(z).
\end{align}
Here, again, these one dimensional FFs  are just equal to the corresponding three dimensional FFs integrated over $d^2k_{F\perp}$ 
with the following four exceptions, 
\begin{align}
& D_{LT}(z) = \int \frac{d^2k_{F\perp}}{(2\pi)^2} \Bigl( D_{LT}(z,k_{F\perp}) + \frac{k_{F\perp}^2}{2M^2} D_{LT}^{\perp}(z,k_{F\perp}) \Bigr), \\
& G_{LT}(z) = \int \frac{d^2k_{F\perp}}{(2\pi)^2} \Bigl( G_{LT}(z,k_{F\perp}) + \frac{k_{F\perp}^2}{2M^2} G_{LT}^{\perp}(z,k_{F\perp}) \Bigr),\\
& H_{1LT}(z) = \int \frac{d^2k_{F\perp}}{(2\pi)^2} \Bigl( H_{1LT}(z,k_{F\perp}) + \frac{k_{F\perp}^2}{2M^2} H_{1LT}^{\perp}(z,k_{F\perp}) \Bigr),\\
& H_{3LT}(z) = \int \frac{d^2k_{F\perp}}{(2\pi)^2} \Bigl( H_{3LT}(z,k_{F\perp}) + \frac{k_{F\perp}^2}{2M^2} H_{3LT}^{\perp}(z,k_{F\perp}) \Bigr).
\end{align}

We list those twist-2 FFs in table \ref{tab:TMDFF1}, and twist-3 in table \ref{tab:TMDFF2}. 
Those at twist-4 have the same structure of those at twist-2, so we will not make a separate table for them.
We also list them according to chiral and time-reversal properties in table \ref{tab:TMDFFChiralTime}.
\end{widetext}

\section{Summary and discussion} \label{sec:IV}
In summary, we presented the results of the general decomposition of quark-quark correlator for fragmentation 
of quark to hadrons with spin 0, 1/2 and $1$ respectively. 
We showed that the correlator is in general expressed as a sum of a spin independent part, 
a vector polarization dependent part and a tensor polarization dependent part. 
For spin-0 hadrons, only the spin independent part is needed, 
for spin-1/2 hadrons, the spin independent part and the vector polarization dependent part are involved, 
while for spin-1 hadrons, all the three parts are necessary. 
The general decomposition leads to totally 72 components of TMD FFs, 8 from spin independent part, 24 from the vector polarization dependent part 
and the other 40 from the tensor polarization dependent part. 
Among them, 18 contribute at leading twist, 36 contribute at twist-3 and the other 18 at twist-4; 
exactly half of them (36) are T-odd, the other half are T-even;  also half are $\chi$-odd and the other half are $\chi$-even. 

These TMD FFs are used in describing a semi-inclusive high energy reaction (see e.g. \cite{Wei:2013csa,Wei:2014pma}). 
We would like to note that usually for a complete description of a semi-inclusive reaction, 
the quark-quark correlator is not sufficient. 
One usually needs quark-$j$-gluon-quark correlator, too ($j=1,2,...$ represents the number of gluons). 
For example, to make a complete calculation up to twist 3, besides the quark-quark correlator discussed here, 
one needs the quark-gluon-quark correlator. These contributions should be taken into account simultaneously.

\section*{Acknowledgements}
This work was supported in part by the National Natural Science Foundation of China
(Nos.11035003 and 11375104),  the Major State Basic Research Development Program in China (No. 2014CB845406) 
and the CAS Center for Excellence in Particle Physics (CCEPP).

\newpage
\begin{widetext}

\begin{table}[!ht]
\caption{The 18 leading twist components of the FFs for quark fragments to spin-1 hadrons}\label{tab:TMDFF1}
\begin{tabular}{p{1.8cm}l@{\hspace{0.6cm}}lp{2.8cm}l} \hline
\begin{minipage}[c]{1.5cm}quark\\ polarization \end{minipage}  & \begin{minipage}[c]{1.5cm} hadron \\ polarization \end{minipage}  
 & TMD FFs  & integrated over $\vec k_{F\perp}$ &  name  \rule[-0.39cm]{0mm}{0.95cm} \\[0.12cm]  \hline 
& \hspace{0.5cm} $U$ & $D_1(z,k_{F\perp})$ &  \hspace{0.5cm} $D_1(z)$  & number density \\ [-0.0cm]\cline{2-5}
       & \hspace{0.5cm} $T$ & $D_{1T}^\perp(z,k_{F\perp})$ & \hspace{0.6cm} $\times$ &  \\ [0.1cm] \cline{2-5}
\hspace{0.5cm} $U$ & \hspace{0.5cm} $LL$ & $D_{1LL}(z,k_{F\perp})$    & \hspace{0.35cm} $D_{1LL}(z)$  & spin alignment \\ [-0.0cm]
      & \hspace{0.5cm} $LT$ & $D_{1LT}^\perp(z,k_{F\perp})$  & \hspace{0.6cm} $\times$ &  \\ [0.1cm]
      & \hspace{0.5cm} $TT$ & $D_{1TT}^\perp(z,k_{F\perp})$  & \hspace{0.6cm} $\times$ &  \\ [0.1cm] \hline
\vspace{0.7cm} \hspace{0.5cm} $L$ & \hspace{0.5cm} $L$  & $G_{1L}(z,k_{F\perp})$ & \hspace{0.5cm} $G_{1L}(z)$ & spin transfer (longitudinal) \\ [-0.8cm]
 & \hspace{0.5cm} $T$ & $G_{1T}^\perp(z,k_{F\perp})$ & \hspace{0.6cm} $\times$ &  \\ [0.1cm] \cline{2-5}
& \hspace{0.5cm} $LT$  & $G_{1LT}^\perp(z,k_{F\perp})$   & \hspace{0.6cm} $\times$ &  \\ [-0.0cm]
 & \hspace{0.5cm} $TT$ & $G_{1TT}^\perp(z,k_{F\perp})$  & \hspace{0.6cm} $\times$ &  \\ [0.1cm] \hline
& \hspace{0.5cm} $U$ & $H_{1}^\perp(z,k_{F\perp})$ &  \hspace{0.6cm} $\times$ & Collins function \\[0.1cm]\cline{2-5}
 & \hspace{0.4cm} $T(\parallel)$ & $H_{1T}(z,k_{F\perp})$ & \vspace{0.15cm} \hspace{0.5cm} $H_{1T}(z)$ & \vspace{-0.2cm} spin transfer (transverse)  \\[0.1cm]
 & \hspace{0.4cm} $T(\perp)$ & $H_{1T}^\perp(z,k_{F\perp})$ &  &   \\[0.1cm]
 \hspace{0.5cm} $T$ & \hspace{0.5cm} $L$ & $H_{1L}^\perp(z,k_{F\perp})$ & \hspace{0.6cm} $\times$ &  \\[0.1cm] \cline{2-5}
 & \hspace{0.5cm} $LL$ & $H_{1LL}^\perp(z,k_{F\perp})$  & \hspace{0.6cm} $\times$ &  \\[-0.0cm]
 & \hspace{0.5cm} $LT$ & $H_{1LT}(z,k_{F\perp}),~H_{1LT}^\perp(z,k_{F\perp})$  & \vspace{-0.2cm} \hspace{0.35cm} $H_{1LT}(z)$ & \vspace{-0.1cm}   \\[0.2cm]
 & \hspace{0.5cm} $TT$ & $H_{1TT}^\perp(z,k_{F\perp}),~H_{1TT}^{\prime\perp}(z,k_{F\perp})$  & \hspace{0.6cm}$\times$, $\times$ &   \\[0.1cm]\hline
\end{tabular}
\end{table}

\begin{table}[!ht]
\caption{The 36 twist-3 components of the FFs for quark fragments to spin-1 hadrons}\label{tab:TMDFF2}
\begin{tabular}{p{1.8cm}l@{\hspace{0.6cm}}lp{2.8cm}l} \hline
\begin{minipage}[c]{1.5cm}quark\\ polarization \end{minipage}  & \begin{minipage}[c]{1.5cm} hadron \\ polarization \end{minipage}  
 & TMD FFs  & \hspace{0.2cm} integrated over $\vec k_{F\perp}$   \rule[-0.39cm]{0mm}{0.95cm} \\[0.12cm]  \hline 
\vspace{1.3cm} \hspace{0.5cm} $U$  & \hspace{0.5cm} $U$ & $E(z,k_{F\perp})$, $D^\perp(z,k_{F\perp})$ &  \hspace{0.6cm} $E(z)$, $\times$   \\ [-1.3cm]\cline{2-5}
       & \hspace{0.5cm} $L$ & $D_{L}^\perp(z,k_{F\perp})$ & \hspace{0.6cm} $\times$   \\ [0.1cm] 
       & \hspace{0.5cm} $T$ & $E_{T}^\perp(z,k_{F\perp})$, $D_T(z,k_{F\perp})$, $D_T^{\perp}(z,k_{F\perp})$ & \hspace{0.6cm} $\times$, $D_T(z)$   \\ [0.1cm] \cline{2-5}
      & \hspace{0.5cm} $LL$ & $E_{LL}(z,k_{F\perp})$, $D_{LL}(z,k_{F\perp})$    & \hspace{0.6cm} $E_{LL}(z)$, $\times$    \\ [0.0cm]
      & \hspace{0.5cm} $LT$ & $E_{LT}^\perp(z,k_{F\perp})$, $D_{LT}(z,k_{F\perp})$, $D_{LT}^\perp(z,k_{F\perp})$  & \hspace{0.6cm} $\times$, $D_{LT}(z)$   \\ [0.1cm]
      & \hspace{0.5cm} $TT$ & $E_{TT}^\perp(z,k_{F\perp})$, $D_{TT}^\perp(z,k_{F\perp})$, $D_{TT}^{\prime\perp}(z,k_{F\perp})$  & \hspace{0.6cm} $\times$, $\times$, $\times$   \\ [0.1cm] \hline
\vspace{1.3cm} \hspace{0.5cm} $L$ & \hspace{0.5cm} $U$  & $G^\perp(z,k_{F\perp})$ & \hspace{0.6cm} $\times$  \\ [-1.3cm] \cline{2-5}
 & \hspace{0.5cm} $L$ & $E_L(z,k_{F\perp})$, $G_L^\perp(z,k_{F\perp})$ & \hspace{0.6cm} $E_L(z)$, $\times$   \\ [0.1cm]
 & \hspace{0.5cm} $T$ & $E_T(z,k_{F\perp})$, $G_T(z,k_{F\perp})$, $G_T^\perp(z,k_{F\perp})$ & \hspace{0.6cm} $\times$, $G_T(z)$   \\ [0.1cm] \cline{2-5}
& \hspace{0.5cm} $LL$  & $G_{LL}^\perp(z,k_{F\perp})$   & \hspace{0.6cm} $\times$   \\ [-0.0cm]
& \hspace{0.5cm} $LT$  & $E_{LT}^{\prime\perp}(z,k_{F\perp})$, $G_{LT}(z,k_{F\perp})$, $G_{LT}^{\perp}(z,k_{F\perp})$   & \hspace{0.6cm} $\times$, $G_{LT}(z)$   \\ [-0.0cm]
 & \hspace{0.5cm} $TT$ & $E_{TT}^{\prime\perp}(z,k_{F\perp})$, $G_{TT}^{\perp}(z,k_{F\perp})$, $G_{TT}^{\prime\perp}(z,k_{F\perp})$  & \hspace{0.6cm} $\times$, $\times$, $\times$   \\ [0.1cm] \hline
& \hspace{0.5cm} $U$ & $H(z,k_{F\perp})$ &  \hspace{0.6cm} $H(z)$  \\[0.1cm]\cline{2-5}
& \hspace{0.5cm} $L$ & $H_L(z,k_{F\perp})$ &  \hspace{0.6cm} $H(z)$  \\[0.1cm]
 & \hspace{0.4cm} $T(\parallel)$ & $H_{T}^\perp(z,k_{F\perp})$ & \hspace{0.6cm} $\times$  \vspace{-0.0cm} \\[0.1cm]
\hspace{0.5cm} $T$ & \hspace{0.4cm} $T(\perp)$ & $H_{T}^{\prime\perp}(z,k_{F\perp})$ & \hspace{0.6cm} $\times$   \\[0.1cm] \cline{2-5}
 & \hspace{0.5cm} $LL$ & $H_{LL}(z,k_{F\perp})$  & \hspace{0.6cm} $H_{LL}(z)$  \\[-0.0cm]
 & \hspace{0.5cm} $LT$ & $H_{LT}^\perp(z,k_{F\perp})$, $H_{LT}^{\prime\perp}(z,k_{F\perp})$  & \hspace{0.6cm} $\times$, $\times$  \vspace{-0.1cm}   \\[0.2cm]
 & \hspace{0.5cm} $TT$ & $H_{TT}^\perp(z,k_{F\perp})$, $H_{TT}^{\prime\perp}(z,k_{F\perp})$  & \hspace{0.6cm}$\times$, $\times$ &   \\[0.1cm]\hline
\end{tabular}
\end{table}

\begin{table}[!ht]
\caption{Chiral and time reversal properties of TMD FFs from quark-quark correlator}\label{tab:TMDFFChiralTime}
\begin{tabular}{@{}ccllll@{}}
\hline
\multirow{2}{*}{\begin{tabular}[c]{@{}c@{}}quark\\ polarization\end{tabular}} & \multirow{2}{*}{\begin{tabular}[c]{@{}c@{}}hadron\\ polarization\end{tabular}}~~~ & \multicolumn{2}{l}{chiral-even} & \multicolumn{2}{l}{chiral-odd} \\ \cline{3-6} 
& & T-even          & T-odd         & T-even         & T-odd         \\ \hline
\multirow{6}{*}{$U$} & $U$  &  $D_1$,~ $D^\perp$,~ $D_3$~ &  & $E$~ &  \\ \cline{2-6}
& $L$ &  & $D_L^\perp$~ &  & \\
& $T$  & & $D_{1T}^\perp$,~ $D_T$,~ $D_T^{\perp}$,~ $D_{3T}^\perp$~ &  & $E_T^\perp$~ \\ \cline{2-6}
& $LL$ & $D_{1LL}$,~ $D_{LL}$,~ $D_{3LL}$~ & & $E_{LL}$~ & \\
& $LT$ & $D_{1LT}^\perp$,~ $D_{LT}$,~ $D_{LT}^\perp$,~ $D_{3LT}^\perp$~ &  & $E_{LT}^\perp$~ & \\
& $TT$ & $D_{1TT}^\perp$,~ $D_{TT}^\perp$,~ $D_{TT}^{\prime\perp}$,~ $D_{3TT}^{\perp}$~~~ & &$E_{TT}^\perp$~ &\\ \hline
\multirow{6}{*}{$L$}& $U$& & $G^\perp$~ &  &   \\ \cline{2-6}
& $L$ & $G_{1L}$,~ $G_L^\perp$,~ $G_{3L}$~ &  &  & $E_L$~  \\
& $T$ & $G_{1T}^\perp$,~ $G_T$,~ $G_T^\perp$,~ $G_{3T}^\perp$~ & &  & $E_T$~  \\ \cline{2-6}
& $LL$ & & $G_{LL}^\perp$~ &  & \\
& $LT$ & & $G_{1LT}^\perp$,~ $G_{LT}$,~ $G_{LT}^\perp$,~ $G_{3LT}^\perp$~ & $E_{LT}^{\prime\perp}$~ & \\
& $TT$ & & $G_{1TT}^\perp$,~ $G_{TT}^\perp$,~ $G_{TT}^{\prime\perp}$,~ $G_{3TT}^\perp$~~~~~ & $E_{TT}^{\prime\perp}$~ & \\ \hline
\multirow{7}{*}{$T$} & $U$ & & & & $H_1^\perp$,~ $H$,~ $H_3^\perp$~ \\ \cline{2-6}
& $L$ & & & $H_{1L}^\perp$,~ $H_L$,~ $H_{3L}^\perp$ & \\
& $T(\parallel)$  &  &  & $H_{1T}$,~ $H_T^\perp$,~ $H_{3T}$ &  \\
& $T(\perp)$ &  & & $H_{1T}^\perp$,~ $H_T^{\prime\perp}$,~ $H_{3T}^\perp$~~~ &  \\ \cline{2-6}
& $LL$ & & &  & $H_{1LL}^\perp$,~ $H_{LL}$,~ $H_{3LL}^\perp$~ \\
& $LT$ &  & & & $H_{1LT}$,~ $H_{1LT}^\perp$,~ $H_{LT}^\perp$,~ $H_{LT}^{\prime\perp}$,~ $H_{3LT}$,~ $H_{3LT}^\perp$~ \\
& $TT$ &  & & & $H_{1TT}^\perp$,~ $H_{1TT}^{\prime\perp}$,~ $H_{TT}^\perp$,~ $H_{TT}^{\prime\perp}$,~ $H_{3TT}^\perp$,~ $H_{3TT}^{\prime\perp}$~ \\ \hline 
\end{tabular}
\end{table}

\end{widetext}

\end{document}